\documentclass[a4paper,twocolumn,11pt,unpublished]{quantumarticle}
\pdfoutput=1
\usepackage[utf8]{inputenc}
\usepackage[english]{babel}
\usepackage[T1]{fontenc}
\usepackage{amsmath,amssymb,amsthm}
\usepackage{hyperref}
\usepackage{cite}
\usepackage[skip=8pt]{parskip}

\usepackage{tikz}
\usepackage{tikz-cd}

\newcommand{\udarr}{
        \begin{tikzpicture}[scale=0.13]          
            \draw [<->] (0, 1) -- (0, -1);        
        \end{tikzpicture}
}
\newcommand{\lrarr}{
        \begin{tikzpicture}[scale=0.15]
            \draw [<->] (-1, 0) -- (1, 0);                
        \end{tikzpicture}
}

\definecolor{QB}{RGB}{133, 199, 187}
\definecolor{dg}{RGB}{34, 177, 76}
\definecolor{lg}{RGB}{165, 209, 27}
\definecolor{or}{RGB}{255, 190, 16}
\definecolor{quantumviolet}{HTML}{53257F}

\hypersetup{colorlinks=true, linkcolor=quantumviolet, citecolor=quantumviolet, urlcolor=quantumviolet}

\newtheorem{definition}{Definition}

\begin{document}

\title{Time-Bin CKA as a tool for blockchain technology}

\author{Marta Misiaszek-Schreyner}
\email{\\misiaszek@quantumblockchains.io}
\orcid{0000-0002-5464-3277}
\affiliation{Quantum Blockchains Inc., Ogrodowa 8, 91-062 Lodz, Poland, \href{http://quantumb.io/}{http://quantumb.io/}}

\author{Miriam Kosik}
\orcid{0000-0002-3660-2199}
\affiliation{Quantum Blockchains Inc., Ogrodowa 8, 91-062 Lodz, Poland, \href{http://quantumb.io/}{http://quantumb.io/}}

\author{Mirek Sopek}
\orcid{0000-0003-0378-5125}
\affiliation{Quantum Blockchains Inc., Ogrodowa 8, 91-062 Lodz, Poland, \href{http://quantumb.io/}{http://quantumb.io/}}

\maketitle

\begin{abstract}
We explore the potential of Time-Bin Conference Key Agreement (TB CKA) protocol as a means to achieve consensus among multiple parties. We provide an explanation of the underlying physical implementation, i.e. TB CKA fundamentals and illustrate how this process can be seen as a natural realization of the \textit{global common coin} primitive. Next, we present how TB CKA could be embodied in classical consensus algorithms to create hybrid classical-quantum solutions to the Byzantine Agreement problem.
\end{abstract}

\section{Blockchain technology}
A blockchain is an architecture that enables data to be stored in a decentralized network \cite{book:Wattenhofer}. The primary distinctions between conventional databases and blockchains encompass decentralization, distribution, the implementation of cryptographic protocols resulting in the linkage of data across blocks, and the inherent immutability of records. \par

In a traditional database, data is stored in a centralized location controlled by a single entity or organization. This creates the need for trust - users must trust this central entity to maintain the data accurately and securely. In a blockchain, data is stored in a decentralized network of computers, referred to as nodes. Each node in the network holds a copy of the entire blockchain and participates in the validation and verification of transactions. There is no central authority, and consensus mechanisms ensure that all nodes agree on the state of the data. \par

Data in a traditional database is typically linked through relationships established between tables using keys. In a blockchain, data is linked through blocks. Each block in a blockchain contains some data and a hash, which is a unique fingerprint that identifies the block. Any change inside the block will cause the hash to change. Each block also contains the hash of the previous block, which leads to a chain of blocks. This linking creates a chronological and immutable transaction history.  \par

The kind of data stored inside a block depends on the type of blockchain (for example, the Bitcoin blockchain stores transaction details, such as the sender, receiver, and number of coins). Blockchain blocks need not necessarily be in the form of uniform binary data blocks. Modern solutions allow for much richer data structures to be linked \cite{Sopek2022} to form the chain. What is essential is that the entire system represents a consistent generalized transaction history on which all nodes achieve eventual agreement about the linked data. \par

The main components of blockchain software are consensus and validation algorithms that provide transparency and data security. Unlike ordinary databases, a public blockchain does not rely on a centralized model of trust because it is fully accessible to anyone who wants to participate as a node. Such node gets a full copy of the blockchain and can even use the copy of the blockchain to verify that everything is in order. Therefore the security of a blockchain comes not only from the creative use of encryption, hashing and consensus mechanisms, but also from being distributed and decentralized.\par

Despite the common features of the blockchain software, there are various consensus mechanisms, for example:  Proof of Work (PoW), Proof of Stake (PoS), Delegated Proof of Stake (DPoS), Proof of Authority (PoA), Proof of Capacity (PoC) and many others \cite{Wang2020}. All of them are mathematical operations through which nodes from the network validate creation of new blocks, however, they differ in the type of algorithm that is used. The most popular and most famous is PoW (used in Bitcoin, early Ethereum and other networks), despite the fact that it needs high computational effort that results in high energy consumption.\par

Despite the numerous advantages of blockchain, it also comes with significant drawbacks stemming from its distributed architecture. In theoretical considerations of distributed systems there are two fundamental theorems that limit desirable properties of blockchain architecture. One of them is known as \textbf{CAP theorem} \cite{Gilbert2012}, the other is \textbf{FLP impossibility result} \cite{FLP}. \par

CAP theorem \cite{Gilbert2012} states that any distributed system can have at most two of the following three properties:
\begin{itemize}
\item consistency (C)  --  every read receives the most recent write;
\item availability (A) --  each request eventually receive a response;
\item partition tolerance (P) -- the system operates despite an arbitrary number of messages being dropped between nodes due to communication breakdowns or any other reasons.
\end{itemize}
Unfortunately, the CAP theorem oversimplifies the balance between these properties. Due to that this formulation is not genuinely true. CAP theorem states only that perfect availability and consistency in the presence of partitions is not possible. Therefore the designers of distributed systems do not need to choose between consistency and availability when partitions are present. The goal is rather to find a trade-off between them. \par

The FLP impossibility result \cite{FLP}, named after its authors (Fischer, Lynch and Patterson), comes from consideration on achieving consensus in distributed systems. It shows that in an asynchronous setting, there is no deterministic distributed algorithm that always solves the consensus problem, even if only one node of the system is fault. \par

The limits ensuing from both CAP and FLP theorems translate to the phenomenon called the \textbf{blockchain trilemma}: \textit{it is impossible for any classical blockchain to simultaneously guarantee security, scalability and decentralization}. Various blockchain consensus algorithms attempted to find a balance between these three features, resembling the trade-offs made by the designers of standard distributed systems. One of the approaches to minimize the negative effects of the trilemma is to prioritize data availability (i.e. scalability) and agree that the data may not be consistent on all nodes at the same time, but to demand that it is eventually consistent, i.e. after some time of the system life. \par 

Since these problems are crucial for the blockchain technology, it is important to analyse new proposals and test new algorithms. Luckily, recent works \cite{Marcozzi2021} show that the use of quantum mechanical laws can be beneficial for the reduction of negative consequences of the trilemma and could lead to an entirely new class of blockchain architectures. \par

However, at the same time the emerging quantum computers pose a threat to the security of modern blockchains, which are built mostly as P2P networks and assume heavy use of the classical asymmetric cryptography with public-private keys playing a pivotal role. Therefore, it is wise to explore the possibilities of integration of quantum cryptography and quantum devices in the blockchain architecture. \par

\subsection{Quantum secured blockchain}
As it was mentioned earlier, quantum computers pose a threat to any classical encryption algorithms that are used nowadays. Therefore, blockchain protection with quantum cryptography is a sensible step in further development of this technology. \par

This development may take many different paths. 
\begin{itemize}
\item One of them is to simply use a quantum random number generator for creating the encryption keys. Since such keys have higher degree of randomness than keys generated using any available algorithms, or obtained using any classical physical processes, this way of communication is way more secure than the communication that is currently provided.

\item The second is the use of quantum key distribution (QKD) devices that are available on the market and setting quantum channels between each node of the blockchain network (one to one architecture). The use of these devices for obtaining consensus significantly increases the security and ensures the validity of a newly created blocks. Also, as shown in Ref.~\cite{Sun2019}, another type of consensus mechanism can be used, which, by using one-time pad encryption keys further improves the security of the blockchain.

\item The third method is to explore different quantum communication protocols that enable the use of other network architectures and favorably affect scalability of such network.

\end{itemize}

The third development path offers the most novelty, therefore, it is elaborated upon further in this work. \par

\section{Quantum Key Distribution}
Similarly to classical data encryption, quantum cryptography is also based on key distribution. The difference is that in quantum cryptography the key is generated by a non-deterministic purely random process. This process, which occurs in accordance with quantum mechanical laws, ensures the security of a key itself and also the security of its distribution between involved parties. For example, since a quantum state collapses when measured, the eavesdropping of transmission can be easily detected. Also, due to the no-cloning theorem \cite{Wooters1982}, it is impossible to copy the data that is encoded in a quantum state. All of this makes quantum key distribution (QKD) \textit{an information-theoretically secure} solution to the key exchange problem. \par

There are many protocols used for QKD. One of the best known is BB84 \cite{BB84}, named after Charles Bennett and Gilles Brassard who presented it in 1984. In this protocol, a secret key is encoded in photons' polarization states, randomly chosen from two available bases. Each photon represents a single bit of data. Its value is established after the transmission of a photon through the quantum channel and the measurement of its polarization state. Since the measurement is done in two bases, that are also randomly chosen, the outcome of the measurements need to be reconciled by communicating parties. It is done through a classical channel (such as phone, mail, HTTP or any similar way of communication). Unfortunately, as a result of the reconciliation procedure, information is leaked and hence, up to a half of the sent bits may need to be removed from the key. Moreover, there are also other losses, decoherence and measurement imperfections that may influence the key generation rate. \par

Furthermore, the standard QKD protocols such as that presented above require to set quantum channels between all communicating parties. It means that for N parties, $\frac{N(N-1)}{2}$ connections are needed. \par

All of this, combined with the high cost of available QKD systems, results in slowdown in development of the commercial use-cases of quantum cryptography. \par

Fortunately, there is a novel QKD protocol that allows to decrease the number of connections in the system to N for N communicating parties. It is called Quantum Conference Key Agreement (CKA). \par

\subsection{Quantum Conference Key Agreement}

\begin{figure}[b!]
\centering
\includegraphics[width=\linewidth]{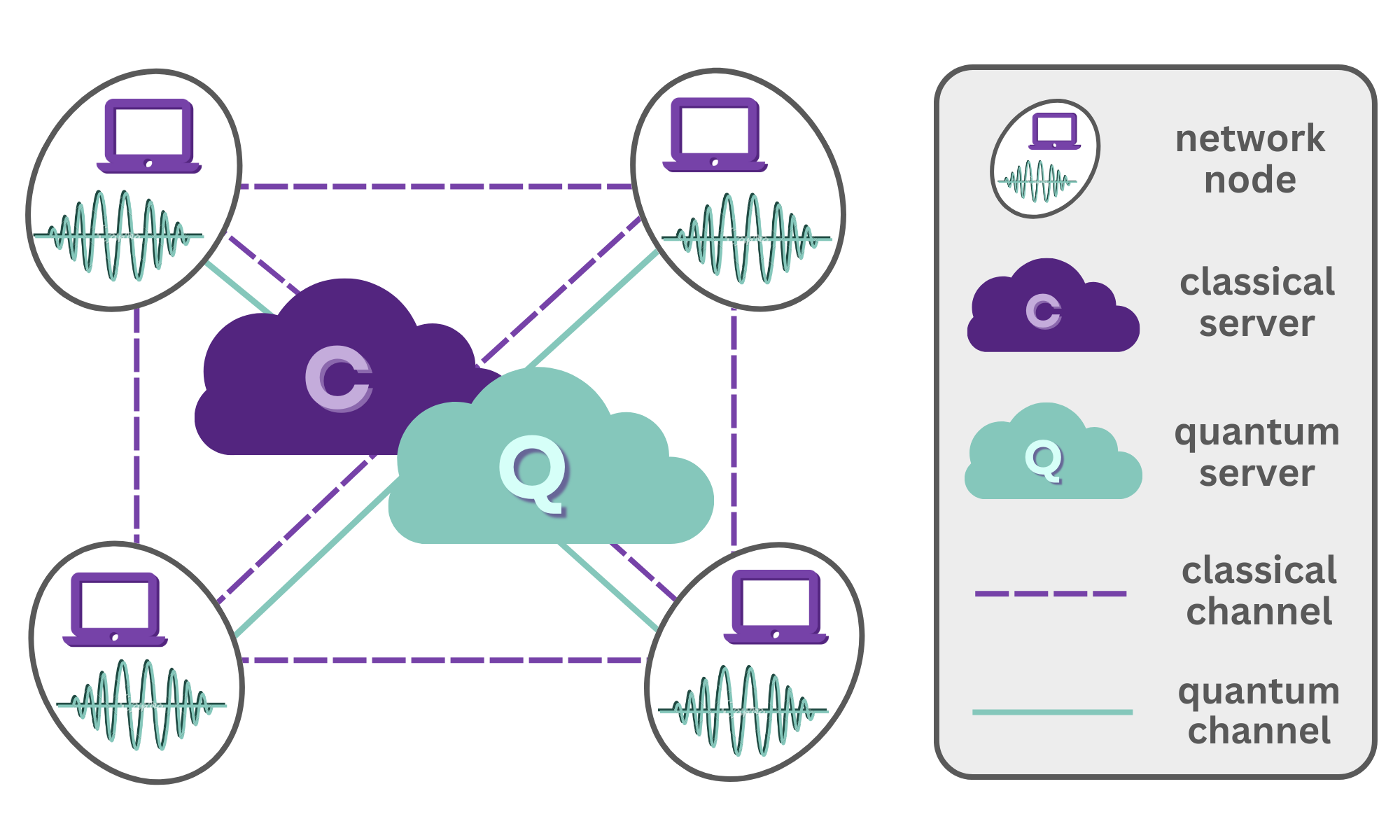}
\caption{Schematical visualisation of a four-node network that allows to realize the CKA protocol. The purple dashed lines indicate classical connections, green solid lines denote quantum connections.}
\label{fig:scheme}
\end{figure}

The Quantum Conference Key Agreement is a protocol that enables multiparty quantum key exchange. Thanks to this protocol, it is possible to exchange a key between many parties in a secure manner. \par

The infrastructure that allows to achieve such consensus consists of two parts. One is a classical Internet cloud that connects all parties using classical communication in classical channels, the other is a quantum server that is responsible for preparing and distributing shared qubits between all parties at once. The scheme of such infrastructure is presented in Fig. \ref{fig:scheme}. \par

CKA is based on sharing N qubits with N communicating parties. These qubits are in a specific entangled state called $|GHZ\rangle$.
The CKA protocol was experimentally demonstrated in a 4-node network in 2021 \cite{Proietti2021}. 
 \par
As it can be seen, the nodes are connected to the quantum server. They are also connected with each other by classical channels, which is not presented in the figure. \par

The quantum server is responsible for the distribution of an entangled state, which is here of the form
\begin{equation*}
|GHZ\rangle=\frac{1}{\sqrt{2}}\Big(|0000\rangle+|1111\rangle \Big) \; .
\end{equation*}
Here, such quantum state is generated using two SPDC sources in Sagnac mode. The single PPLN crystal, pumped by pulsed Ti:Sapphire laser generates a quantum state that can be written in the form of
\begin{equation}
|\psi\rangle=\frac{1}{\sqrt{2}}\big(|0\rangle+|1\rangle\big) \; ,
\end{equation}  
where $|0\rangle$ and $|1\rangle$ are orthogonal polarization states. Generated qubits are correlated with each other using PBS (polarization beamsplitter) and distributed using long single-mode fibers. Then, each node measures its qubit similarly to the method used for standard BB84 protocol (collecting detections for different settings of quarter and half waveplates). \par


On the one hand, although the experimental realization was presented only for 4 nodes separated from each other by 50 km at maximum, due to the scalability of this method, it is a promising solution worth to consider in further development of quantum consensus mechanism in distributed systems. On the other hand, due to decoherence, encoding information in the means of photon polarization state is not the best choice for long-range communication using optical fiber links. Therefore, other implementation of CKA protocol should be addressed.\par

\subsection{Time-bin CKA}

\begin{figure}[t!]
\centering
\includegraphics[width=0.55\linewidth]{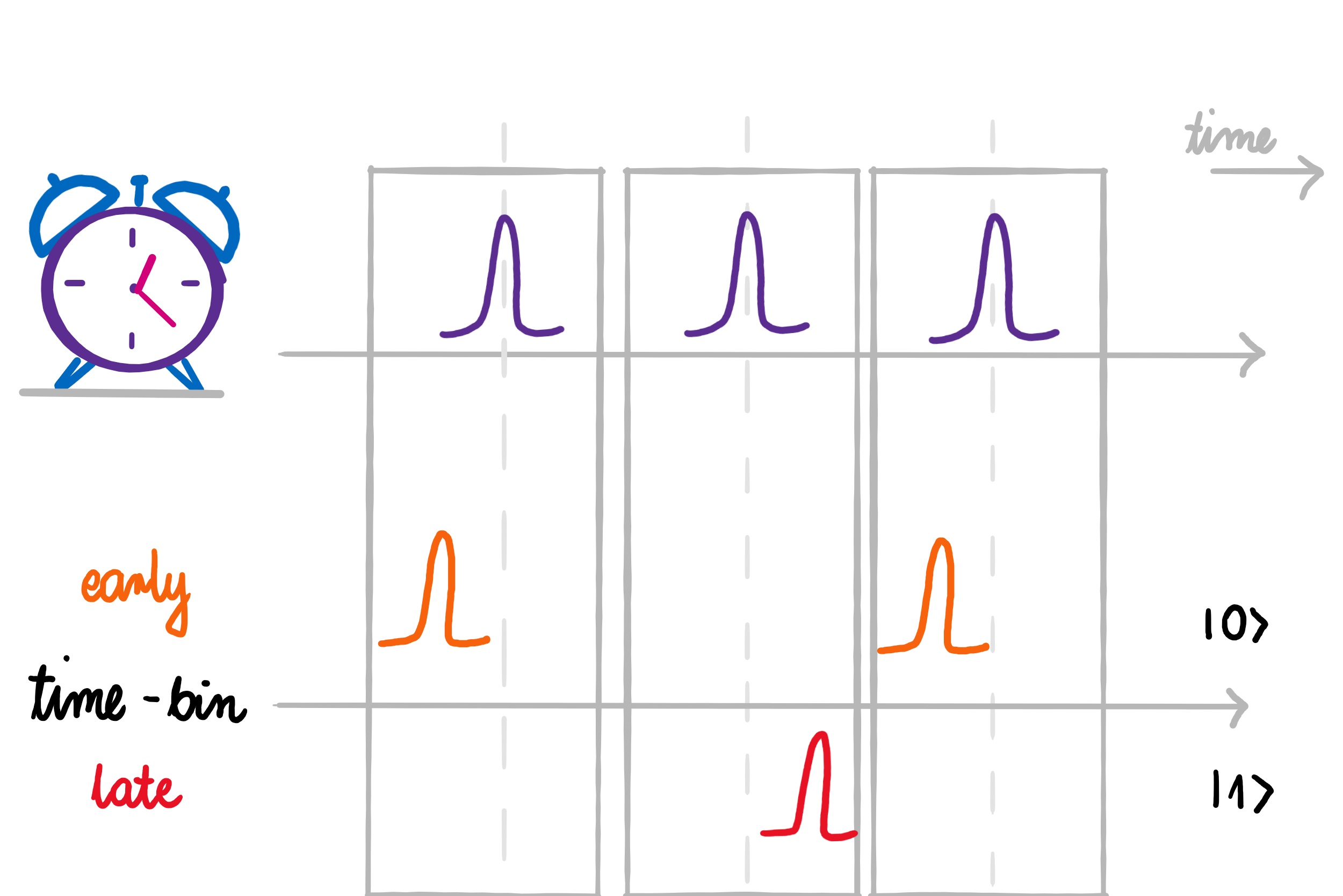}
\caption{Scheme of time-bin encoding. Early- and late- time-bin in relation to the external clock. The illustration has been sourced from Ref. \cite{Misiaszek2022} with the author's explicit consent.}
\label{pic:tbe}
\end{figure}

One of them could be the use of time-bin encoding. Such qubits may be created in a simple way by single photons traveling along paths of different lengths. As a result, photons' arrival time, compared to external clock, differ and may be assigned as the early- and late-time-bin, as presented in Figure \ref{pic:tbe}. \par

\begin{figure}[t!]
\centering
\includegraphics[width=0.9\linewidth]{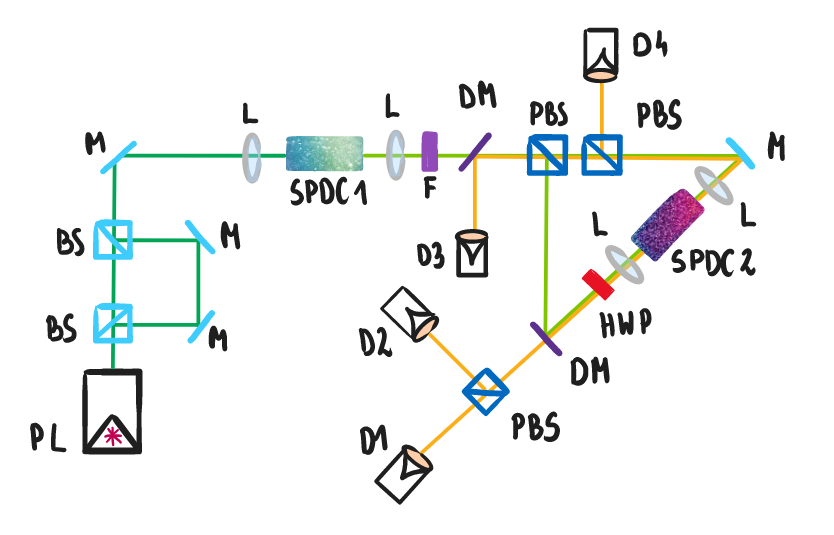}
\caption{Proposal of the experimental setup for implementation of TB CKA protocol. Symbols: PL -- pulsed laser, BS -- beamsplitter, M --mirror, L -- lens, F -- filter, DM -- dichroic mirror, PBS -- polarization beamsplitter, SPDC1, SPDC2 -- nonlinear crystals, D1,D2,D3,D4 -- single-photon detectors. Converted photon beams are marked with different color lines.}
\label{pic:ckatb}
\end{figure}

Figure \ref{pic:ckatb} shows the proposal of experimental setup that enables to introduce CKA protocol with time-bin encoding for four node network. \par
Let us follow a beam path inside the setup. 
At the beginning, the beam from a pulsed pump laser (PL) enters a set of beamsplitters and mirrors, resulting in creation of time bins, that may be denoted as "states" $|0\rangle$ and $|1\rangle$ (\textcolor{dg}{\textbf{dark green}} line). Then, the beam is focused on first SPDC nonlinear crystal (SPDC1), which creates a polarization qubit of the form
\begin{equation}
|\phi\rangle=\frac{1}{\sqrt{2}}\big(|\udarr\rangle+|\lrarr\rangle\big)\; ,
\end{equation}
where $|\udarr\rangle$, $|\lrarr\rangle$ are vertical and horizontal polarization states, respectively. Taking into account previously created time bins, at this point the photon state from the beam (marked with \textcolor{lg}{\textbf{light green}} line) can be described as
\begin{equation}
|\Phi\rangle=\frac{1}{\sqrt{2}}\big(|0_{\udarr}0_{\lrarr}\rangle+|1_{\udarr}1_{\lrarr}\rangle\big)\;.
\end{equation}
Later, the \textcolor{lg}{\textbf{light green}} photon beam is directed onto a polarization beamsplitter (PBS), that divides the  beam into two, that are used to build the Sagnac single-photon source (in which beams pass through the crystal in opposite directions simultaneously). Let us assume that in described case beam with horizontal polarization ($\lrarr$) passes through PBS. Then, this beam is focused into the second nonlinear crystal (SPDC2), where another pair of orthogonally polarized photons are generated, so the \textcolor{lg}{\textbf{light green}} beam here is converted into \textcolor{or}{\textbf{orange}} one in such a way:
\begin{equation}
\big(|0_{\lrarr}\rangle+|1_{\lrarr}\rangle\big) \,\big(|0_{\udarr}0_{\lrarr}\rangle+|1_{\udarr}1_{\lrarr}\rangle\big) .
\end{equation}
Then, the \textcolor{or}{\textbf{orange}} beam passes through half-wave plate that rotates the polarization state of each photon, resulting in a similar state $\big(|0_{\lrarr}0_{\udarr}\rangle +|1_{\lrarr}1_{\udarr}\rangle\big)$ . This photon beam is then separated by the dichroic mirror (DM), passes through another PBS and are detected in detectors D1 and D2. \par
In the other arm, beam with vertical polarization ($\udarr$) is reflected from PBS, and passes through HWP that changes its polarization to horizontal one ($\lrarr$). It enables to convert photon beam (\textcolor{lg}{\textbf{light green}} line) in the same manner for both beams that create the Sagnac source arms, resulting in a state  $\big(|0_{\lrarr}0_{\udarr}\rangle+|1_{\lrarr}1_{\udarr}\rangle\big)$ (\textcolor{or}{\textbf{orange}} line). Later, the beams that differ with polarization are separated in PBS and detected in D3 and D4. It should be noted here that lines with different colors mark each state of beam conversion (wavelength change). \par

Finally, taking into account both conversion processes, the overall photon state before the measurement can be written as
\begin{equation}
|\Psi\rangle=\frac{1}{\sqrt{2}}\big(|0_{\lrarr}0_{\udarr}0_{\lrarr}0_{\udarr}\rangle+|1_{\lrarr}1_{\udarr}1_{\lrarr}1_{\udarr}\rangle\big) \; .
\end{equation}
Such photon state shows so-called hyperentanglement, where in correlation between photons various properties (degrees of freedom) are engaged. \par

Let us also check the kind of photon states that are prepared for detection by respective detectors:
\begin{equation}
\begin{split}
D1:& \; \; |0_{\lrarr}\rangle+|1_{\lrarr}\rangle \;,\\
D2:& \; \; |0_{\udarr}\rangle+|1_{\udarr}\rangle\;,\\
D3:& \; \; |0_{\lrarr}\rangle+|1_{\lrarr}\rangle\;,\\
D4:& \; \; |0_{\udarr}\rangle+|1_{\udarr}\rangle\;.
\end{split}
\end{equation}

Since single-photon detectors are not usually sensitive to the photon polarization state (they are more sensitive to photon wavelength), it may be concluded that in all detectors the same shared time-bin qubit is measured. Thanks to this it is possible to establish the key simultaneously with all four nodes. \par

\section{Quantum distributed consensus algorithm and FLP impossibility}
\label{sec:protocol}

As it was mentioned earlier, the \textbf{GHZ state} is one of the quantum mechanical tools that enables to achieve distributed consensus \cite{Marcozzi2021, Hondt2004}. In general, GHZ state is an ensemble of N entangled qubits, whose state may be mathematically written as
\begin{equation*}
|GHZ\rangle=\frac{1}{\sqrt{2}}\Big(|0\rangle^{\otimes N} + |1\rangle^{\otimes N} \Big) \; .
\end{equation*}
Similarly to the experiment present in previous section, each node in a network receives a single qubit and measures it, choosing "0" if measured state is $|0\rangle$ and 1 otherwise. Due to that, the single measurement causes collapse of qubit state to $|0\rangle$ or $|1\rangle$ for each participant of communication, not only those who made the measurement, this allows for obtaining a consensus on single bit of information between multiple nodes. \par

Scheme of a protocol that allows to obtain consensus over a proposed block of data $d_1$ may look as follows:
\begin{enumerate}
\item the CKA protocol provides each participating node with two random bits: $b_1$ and $b_2$; if the CKA protocol succeeded without errors, then every participating node should have the same value of b1 and b2 as other nodes;
\item every node performs the calculation: $b_1\; \text{XOR}\; b_2 = b_3$; this step ensures that in later communication no-one will expose the bare value of $b_1$ or $b_2$;
\item every node shares $b_3$ with all other nodes using an authenticated classical communication channel
    \begin{itemize}
    \item nodes with the same value of $b_3$ (the majority) perform an operation on the data block $d_1$:  $b_1 \; \text{XOR} \;d_1 = d_2$ 
    \item nodes with different value of $b_3$ do nothing (they are temporarily excluded from adding data to the blockchain); 
 \end{itemize}
\item every node shares $d_2$ with others, again \textit{via} the classical authenticated channel
\item every node waits until it has received the value of $d_2$ from all other nodes; if all the values of $d_2$ which it has received are identical, then it accepts the block $d_1$ as a new block in the blockchain; 
\end{enumerate}

It should be noted here, that only the first step of a protocol requires access to the quantum channel (bits $b_1$ and $b_2$ come from the measurement of a quantum state that is shared between parties). All other steps are performed using classical communication layers. \par

The presented method of obtaining consensus provides all properties of distributed consensus \cite{Wang2020, FLP}:
\begin{itemize}
\item agreement -- provided by quantum mechanics (measurement of any entangled qubit cause all other qubits to collapse into an identical state);
\item validity -- provided by proposing either "0" or "1" after the measurement done by first node;
\item wait-free processing -- provided by the entanglement, i.e. the quantum state of all qubits collapses simultaneously.
\end{itemize}

It is worth to mention that faulty nodes do not influence achievement of the consensus, because the consensus is obtained after any measurement performed by any node. \par

Summarizing, the use of GHZ state enables to achieve the consensus in distributed systems and, what is more, overcome FLP impossibility result. \par

\section{CKA as a common coin for randomized protocols}
\label{sec:randomized_consensus}

As stated before, reaching consensus by means of a deterministic protocol is impossible~\cite{FLP}. However, several randomized protocols have been invented, which allow to overcome the limitations of FLP~\cite{Rabin1983, BenOr1983, Dwork1990, Chor1989}. The new element in randomized protocols is that they require nodes to perform a coin-toss as an operation. In other words, they assume that each node has access to a random number generator. \par

This novelty, no matter how useful, also constitutes a big practical challenge. The first problem is that in a randomized protocol, all participating nodes require access to perfect randomness. In a classical setting, this requirement on its own is already a big problem. However, there is more. Randomized consensus protocols require a much stronger notion, namely that all participating nodes have simultaneous access to \textit{shared} random bits. \par

The concept of a random bit which is shared among many parties is known as the \textit{common coin}~\cite{Rabin1983, BenOr2005}. 

\begin{definition}[Ref.~\cite{BenOr2005}, p. 3]
Let $G$ be a protocol for $n$ players (with no
input) where each player $P_i$ outputs a (classical) bit $v_i \in \{0, 1\}$. We say that the protocol $G$ is a \textbf{\textit{t}-resilient common coin} protocol with fairness $p$ > 0, if in a system with no more than $t$ faulty nodes $v_i = b$  for any value $b \in \{0, 1\}$, with probability at least $p$, for all good players $P_i$. \par

A common coin with fairness $1/2$ is called a \textbf{strong} common coin.
\end{definition}

The importance of the common coin primitive stems from the fact that the ability to establish a $t$-resilient weak common coin between many parties immediately implies the existence of a $t$-resilient Byzantine Agreement protocol~(Theorem 2 in Ref.~\cite{Canetti1993}). \par

Recent results by Ittai \textit{et al.}~\cite{Ittai2022} state even stronger facts - not only a common coin implies the existence of any consensus protocol but the resulting protocols are time efficient. Consider the problem of asynchronous Binary Agreement with adaptive security, optimal resilience, asymptotically optimal message complexity for a network of N parties, where $t < \frac{N}{2}$ parties may be faulty. Then, given a strong $t$-resilient common coin, there exists a protocol that reaches termination in 7 communication rounds in expectation (Theorem 1.1. in Ref.~\cite{Ittai2022} ). \par

However, the creation of a shared common coin (especially a strong one!) is not a trivial task. Classical methods include the use of Shamir's secret sharing scheme~\cite{Rabin1983} or a Verifiable Secret Sharing Scheme~\cite{Canetti1993, Cachin2002}. These are complicated protocols and moreover they require an additional randomness source to guarantee perfect fairness of the resulting coin. \par

Quantum physics provides a natural solution to this problem. The quantum CKA protocol is a straightforward realization of the \textit{common coin} concept. Moreover, due to the inherent randomness of quantum mechanics it immediately provides the strong version of a common coin.

\section{Other quantum protocols}
\label{sec:quantum_consensus}

There have been several notable ideas for quantum consensus protocols in the recent years~\cite{BenOr2005, Gaertner2008, Rahaman2015, Luo2019, Cholvi2022}. Below, we briefly outline their main ideas, indicating the differences between the existing works and our approach.

\begin{itemize}
    \item Ben-Or, Hassidim (2005)~\cite{BenOr2005}: This was the first quantum approach to the consensus problem. It requires all-to-all communication, which may not be practical for large-scale systems. Also, the paper does not discuss the practical realization of the proposed method.
    \item Rahaman, Wiesniak, Żukowski (2015)~\cite{Rahaman2015}: This interesting paper provides a solution for the pure version of the Byzantine Agreement problem. However, the protocol is only presented for three parties and lacks generalization for more participants. Moreover, there is no mention of an experimental realization of the proposed solution.
    \item Luo, Feng, Zheng (2019)~\cite{Luo2019}: This approach uses $d$-dimensionally entangled states to reach consensus. The list of states to entangle is defined before the protocol begins. When new nodes are added, longer lists need to be generated, potentially impacting the scalability of the method. It is capable of achieving detectable Byzantine Agreement, meaning that an abort action may be required in some steps. The protocol works for up to $N$/3 dishonest parties but, again, the paper lacks details on practical realization. Moreover, it requires the use of a third trusted party (quantum server).
    \item Cholvi (2022)~\cite{Cholvi2022}: 
    Similarly to the previous paper, this approach achieves detectable Byzantine Agreement using Q-correlated lists. However, this protocol works for arbitrary number of dishonest parties, which is a strong result. However, it also lacks details on the practical realization of the method and involves the use of a quantum TTP or quantum server.
\end{itemize}

In summary, the works outlined above focus primarily on theoretical aspects of reaching consensus but tend to overlook experimental realization details. The experimental part is covered in~\cite{Gaertner2008}, this paper however only describes a protocol for three parties and offers no generalization to a setting with more participants. Our work, on the other hand, aims to address both the theoretical and experimental realization aspects to bridge the gap between theory and practical implementation in achieving consensus. \par

\section{Further work and summary}

All information presented above strongly suggests that the advancement of blockchain technology can greatly benefit from leveraging the laws of quantum mechanics. Therefore, it becomes important not only to switch the type of communication from classical to quantum but also to investigate novel models of consensus algorithms, which reflect the unique challenges posed by quantum networks' topologies and architectures. \par

In light of this, although certain amendments are necessary for its application in commercial products, the CKA protocol emerges as an interesting solution to the consensus problem in distributed systems. This is particularly true if it has the potential to challenge the long-standing FLP impossibility result, as demonstrated above. \par

It is important to note that, similarly to some works mentioned in Section~\ref{sec:quantum_consensus}, our solution requires the use of a quantum server, meaning it requires some trust. In the future, we plan to explore the possibility to reformulate our approach in a way where the generation of entanglement can be performed by the participating nodes themselves (e.g. in each round a randomly chosen node in the network acts as the quantum server). This would lead to a truly decentralized consensus protocol. \par 

Moreover, considering the recent advancements in developing quantum networks~\cite{Earl2022} and simulators \cite{DiAdamo2003,Coopmans2010}, we aim to adapt our protocol in a way that would make it suitable for testing within such frameworks. This would allow us to attain practical validation of its efficacy. \par

\subsection*{Author contributions}
MMS conceived the general idea for the manuscript, formulated the main problem. MS contributed to the development of the concept within the classical part and acquired funding for the research. MMS authored Sections 1, 2 and 3, formulated the protocol in Section~\ref{sec:protocol} and created Figures~\ref{pic:tbe} and ~\ref{pic:ckatb}. MK authored Sections~\ref{sec:randomized_consensus} and~\ref{sec:quantum_consensus}, and created Figure~\ref{fig:scheme}. All authors reviewed and validated the entire manuscript, and approved the submission of the manuscript.

\subsection*{Acknowledgements}
Research reported in this paper was partially funded by the Polish Agency for Enterprise Development, grant PARP-POPW.01.01.02-06-0031/21, and by the National Centre for Research and Development (Action 1.3/1.3.1) as part of the Bridge Alfa investment project managed by LT Capital VC fund. We wish to thank our grantors and investors for their support.

\bibliographystyle{quantum}

\end{document}